# Tracing d-d transitions in FePS$_3$ on ultrafast time scales

Jonah Elias Nitschke[1], Michael Gutnikov[1], Karl Schiller[1], Eugenio Coronado[2], Alan Omar[3], Giovanni Zamborlini[1], Clara Saraceno[3], Matija Stupar[1], Alberto M. Ruiz[2], Dorye L. Esteras[2], José J. Baldoví[2], Frithjof Anders[1], Mirko Cinchetti[1]*

[1]TU Dortmund University; Otto-Hahn-Straße 4, 44227 Dortmund, Germany.
[2]Instituto de Ciencia Molecular (ICMol), Universidad de Valencia, Spain.
[3]Photonics and Ultrafast Laser Science (PULS); RU Bochum, Germany.

*Email: mirko.cinchetti@tu-dortmund.de

**Excitations between localized 3d states of transition metal ions within crystalline solids, commonly known as d-d transitions, play a pivotal role in diverse phenomena across solid-state physics, materials science, and chemistry. These transitions contribute to the coloration in transition metal oxides, catalytic processes on oxide surfaces, and high-temperature superconductivity. They also couple optical excitation to quantized collective phenomena such as phonons and magnons in magnetic systems. Until now, an experimental method to unravel the complex quasiparticle dynamics associated with d-d transitions has remained elusive. We bridge this gap by demonstrating that d-d transitions can be distinctly traced in momentum space and time using time- and angle-resolved photoelectron spectroscopy (trARPES). Through this approach, we can assign specific momentum-dependent characteristics and elucidate the decay mechanisms of specific d-d transitions in FePS$_3$, a two-dimensional van der Waals antiferromagnet with a rich array of quantum phenomena stemming from d-d transitions. This study pioneers the use of ARPES in probing the dynamics of d-d transitions across a wide spectrum of solid-state systems.**

# 1. Introduction

Excitations between 3d states of transition metal ions within crystalline solids significantly influence the optical and electronic properties of materials. For instance, they often determine the color of transition metal (TM) oxides[1] and are pivotal in catalytic processes on oxide surfaces [2]. In Cu-O-based perovskites, the interaction between localized Cu 3d holes and more delocalized O 2p holes, leading to a formation of a Zhang-Rice singlet[3], is fundamental for high-temperature superconductivity[4]. In the realm of van der Waals magnets, particularly transition metal phosphorus trisulfides (TMPS$_3$), the transition metal electrons are predominantly localized at the transition metal sites. This localization suggests a similar phenomenology related to d-d transitions as observed in TM oxides. Importantly, when the TM equals Fe, Co, Ni, or Mn, these compounds exhibit behavior characteristic of two-dimensional antiferromagnets. The magnetic moments are primarily localized on the TM ions (Mn$^{2+}$, Fe$^{2+}$, Co$^{2+}$, Ni$^{2+}$) and arise from the interplay between crystal field splitting and exchange splitting. It is plausible to anticipate that an excitation of the ($d^5$, $d^6$, $d^7$, $d^8$) multiplets would be coupled to magnetic excitations as well as to lattice distortions, as the d-d transitions are known to induce static and dynamic Jahn-Teller effects[5].

Recent studies have indeed unveiled that in the XY-type antiferromagnet NiPS$_3$, d-d transitions can generate electron-phonon bound states[6] and selectively activate magnon modes with sub-THz frequency[7]. Coherent excitonic excitations based on Zhang-Rice states, consisting of a hole spin in a localized Ni 3d orbital and a hole spin delocalized over the 3p orbitals of the adjacent ligands, have also been documented in NiPS$_3$[8]. These spin-orbit entangled excitons were demonstrated to be intrinsically coupled to the antiferromagnetic order[8,9], and later used to generate a novel transient conducting antiferromagnetic state[10]. In the Ising-type antiferromagnet FePS$_3$, the excitation of a d-d transition of the Fe$^{2+}$ multiplet has been shown to generate coherent THz optical lattice and hybridized phonon-magnon modes[11], a phenomenon aligning with the strong coupling between antiferromagnetic magnons, and optical phonons reported in 2D antiferromagnetic materials[12–14].

Despite their pivotal role, the dynamics and lifetimes of d-d transitions remain largely unexplored. Understanding these aspects is crucial for harnessing d-d transitions to manipulate the functional properties of 2D quantum materials. Early investigations employing optical absorption spectroscopy[15,16]and x-ray photoelectron spectroscopy[17] elucidated the spectral features arising below the band gap in various TMPS$_3$ crystals (NiPS$_3$, FePS$_3$, ZnPS$_3$, MnPS$_3$) and attributed them to transitions within the d-electron manifold of the corresponding TM atom in octahedral symmetry. These findings facilitated the determination of the crystal field D$_q$ and the Racah interelectron repulsion parameter B, establishing these materials as ionic compounds with localized d-states.

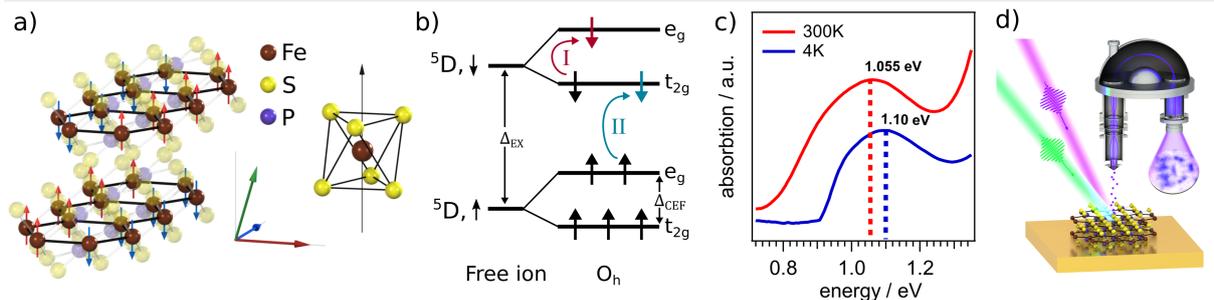

**Figure 1. Sample properties and trARPES setup.** (a) Crystal structure of FePS$_3$ showing the iron-atoms (brown) in a hexagonal structure surrounded by sulfur (yellow) and phosphor (blue) bipyramidals. The octahedral arrangement of the surrounding sulfur atoms leads to the crystal field splitting depicted in (b). The d-orbitals split in the occupied t$_{2g}$ and unoccupied e$_g$ orbitals with the first multiplet d-d transition indicated at 1.08 eV (I). The second possible transition (II) involves a spin flip with an excitation energy of 1.79 eV. (c) Absorption spectrum of FePS$_3$ below the band gap for T=300K and T = 4K, unveiling the slight blue-shift of the d-d transition maximum with decreasing temperature. (d) Schematic visualization of the trARPES experiments.

In this work, we delve into FePS$_3$, aiming to introduce multidimensional time- and angle-resolved photoelectron spectroscopy as a potent experimental technique to bridge the existing knowledge gap concerning the transient dynamics of d-d transitions. FePS$_3$ is a van der Waals antiferromagnetic semiconductor with a band gap of approximately 1.5 eV[18–20] and its electronic properties can be elucidated, to a first approximation, using an ionic model [17]. This model features covalently bonded $(PS_3)_2^{2-}$ clusters exhibiting octahedral symmetry, and Fe$^{2+}$ ions (See **Figure 1a**), whose localized 3d levels are affected by strong electronic correlations. Given that the top of the valence band and the bottom of the conduction band are formed mainly by the Fe(3d) states[21], FePS$_3$ can be classified as a Mott-Hubbard type insulator[18]. Consequently, its low-energy electronic configuration is primarily governed by the Fe(3d) states[20].

The ground state of a Fe$^{2+}$ free ion (a d$^6$ system) is denoted as $^5$D. In FePS$_3$, the octahedral crystal field generated by the $(PS_3)_2^{2-}$ clusters splits this $^5$D state into $^5$T$_{2g}$ and $^5$E$_g$ states, with the latter being higher in energy. The transition between these two states ($^5$T$_{2g}$→$^5$E$_g$) is the only spin allowed transition within the crystal field-split Fe$^{2+}$ multiplet (see I in **Figure 1b**). In contrast, the other possible transitions within this multiplet, as illustrated by II in **Figure 1b**, are spin forbidden and have been rarely studied because of their weak nature. Depending on their energy, these transitions appear in the optical absorption spectra either as a very weak and broad feature centered around approximately 1.1eV (**Figure 1c**) or are masked by the enhanced absorption occurring with the onset of the band gap at around 1.5eV. This is the case both for the spectra recorded at room temperature (300K) as well as at 4K, in agreement with literature [16].

## 2. Results

### 2.1 Below band gap excitation

To highlight the capability of trARPES in capturing the intricated ultrafast dynamics associated to intra-ionic d-d transitions in FePS$_3$, we designed experiments around two selected photon energies: hv$_1$ = 1.2 eV and hv$_2$ = 2.4 eV. These energies are strategically chosen, as they lie below and above the band gap of FePS$_3$, respectively. By using these two photon energies, we optically excite FePS$_3$ and study the subsequent d-d transition dynamics via photoemission with extreme ultraviolet (XUV) probe pulses (**Figure 1d**). The XUV radiation is linearly p-polarized with a photon energy of 21.6 eV, impinging on the sample at an angle of 68° to the surface normal. The cross-correlation of pump and probe pulse (i.e. the temporal resolution of the setup) is around 50 fs.

We start reporting the experiments performed with excitation energy hv$_1$ = 1.2 eV, which is below the band gap and in resonance with the spin-allowed $^5T_{2g} \rightarrow ^5E_g$ transition. **Figure 2a** shows the angle-integrated transient photoemission signal. For consistency, throughout this study, the energy of the photoelectrons is always referred to the valence band maximum (VBM) extracted from static measurements. The data uncovers a transient signal that extends from the VBM to 0.9 eV above it. Remarkably, the signal decays on a timescale of a few-hundred femtoseconds and retains consistent time-dependent characteristics across all energy levels. In **Figure 2b** we show the signal integrated between 0.3eV and 0.6eV above the VBM together with a fit to the data performed with a Gaussian convoluted with a single exponential decay. The decay constant extracted from the fit is $\tau_{blue}$ = 83 ± 10 fs.

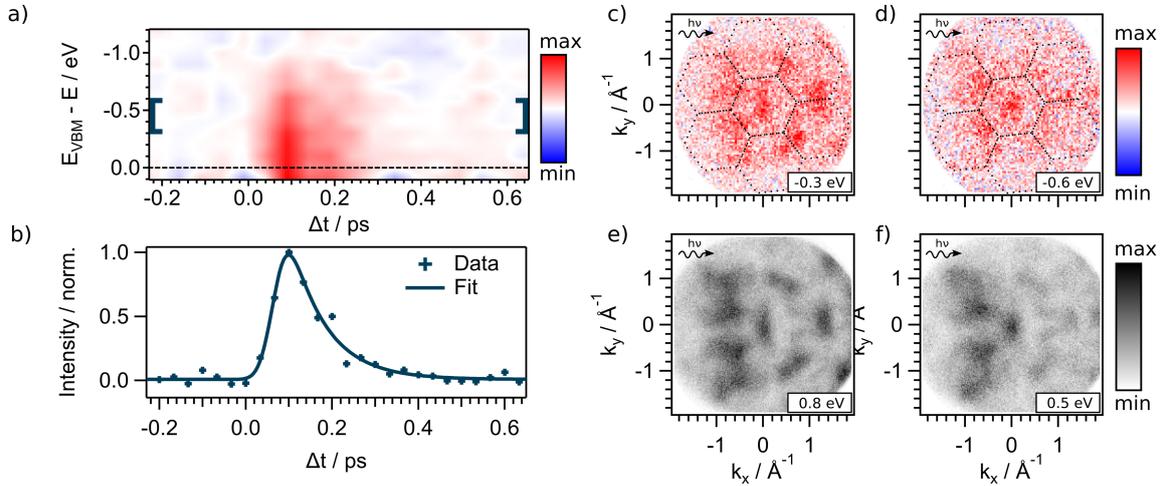

**Figure 2. Visualization and analysis of dynamics for hv$_1$=1.2eV excitation.** (a) Differential angle-integrated transient signal following optical excitation with 1.2 eV pump photons, interpolated with a bilinear algorithm. The depicted difference is between the measured signal and an average from a 200 fs window prior to t$_0$=0 fs. The color gradient spans from blue (negative difference) to red (positive difference), with white indicating no change. A prominent transient signal is observed up to 1 eV above the VBM, though noise increases below the VBM. The blue bracket highlights the region of interest (ROI) used for subsequent analysis. (b) Normalized photoemission intensity of the transient signal, derived by integrating between 0.3 and 0.6 eV above the VBM. (c-d) Differential momentum cuts around t$_0$ for 0.3 eV and 0.6 eV above the valence band maximum (VBM), respectively. A distinct k-dependent pattern is evident, vanishing within hundreds of fs. (e-f) Static momentum maps for 0.8 eV and 0.5 eV below the VBM, respectively.

With the aid of our advanced multi-dimensional time- and angle-resolved photoelectron setup, we can report the transient momentum-resolved patterns captured 130fs post-optical excitation. These differential momentum maps (elaborated further in **Supplementary Fig. S3 and Fig. S5**) are visualized in **Figure 2c and 2d** for energy levels 0.3 eV and 0.6 eV above the VBM. Due to the high angular acceptance of our momentum microscope (± 90°) and the large unit cell of FePS$_3$, we can observe the signal stemming from the first as well as from the neighboring Brillouin zones (BZ), indicated by the black hexagons. The obtained momentum resolved signal unveils a complex fingerprint with higher intensities at the Γ-point of the 1$^{st}$ BZ and the M-points of the 1st and neighboring BZ. The most striking characteristic is their strong similarity to the momentum maps of occupied states, specifically the bands close to the VBM, that stem mainly from the Fe d-orbitals. In fact, the transient signals closely mirror the occupied states around 1.1 eV lower in energy, almost equivalent to the excitation energy of $h\nu_1$ = 1.2 eV, as depicted in **Figures 2e and 2f**. Given the asymmetric nature of the observed transient signal around $t_0$, and its presence even under s-polarized pump light, we rule out laser-assisted photoemission[22] as its cause (details in **Supplementary Fig. S2 and Fig. S6**). Instead, we attribute this unique fingerprint to the $^5T_{2g} \rightarrow {}^5E_g$ transition, with its electronic components being captured by the XUV probe pulse. More insights on this are provided in the discussion section below.

## 2.2 Above band gap excitation

When we examine excitation with a pump energy of $h\nu_2$ = 2.4 eV, the results differ significantly from the earlier discussed pump energy of $h\nu_1$ = 1.2 eV. Specifically, an excitation above the band gap using $h\nu_2$ = 2.4 eV presents a more intricate scenario. **Figure 3a** presents the angle-integrated transient photoemission, uncovering the observed dynamics for unoccupied states up to 2.4 eV above the VBM. In contrast to the observations from the sub-band gap excitation, unique transients emerge across different energy regions. Color-coded brackets (purple, red, dark blue and yellow) demarcate these energy regions, with the derived normalized photoemission intensity transients shown in **Figure 3b**. The highest unoccupied states, indicated by the yellow region around -1.9eV, manifest a single exponential decay with a time-constant of $\tau_{yellow}$ = 115 ± 100 fs. This rapid decay is evident across all energies above the conduction band minimum (CBM), which is located around 1.5 eV above the VBM, according to literature[19]. Venturing slightly below the CBM, we observe a roughly 600 meV broad window centered around 1.1 eV above the VBM that exhibits a double exponential decay. For a closer investigation of these two components, the transient signal around -1.1 eV was measured again in a separate measurement with a 3 times smaller step size of 33 fs (red curve in **Figure 3b**). This unveils a fast decay constant of $\tau_{red,1}$ = 46 ± 34 fs and a long-lived contribution with $\tau_{red,2}$ = 1273 ± 268 fs.

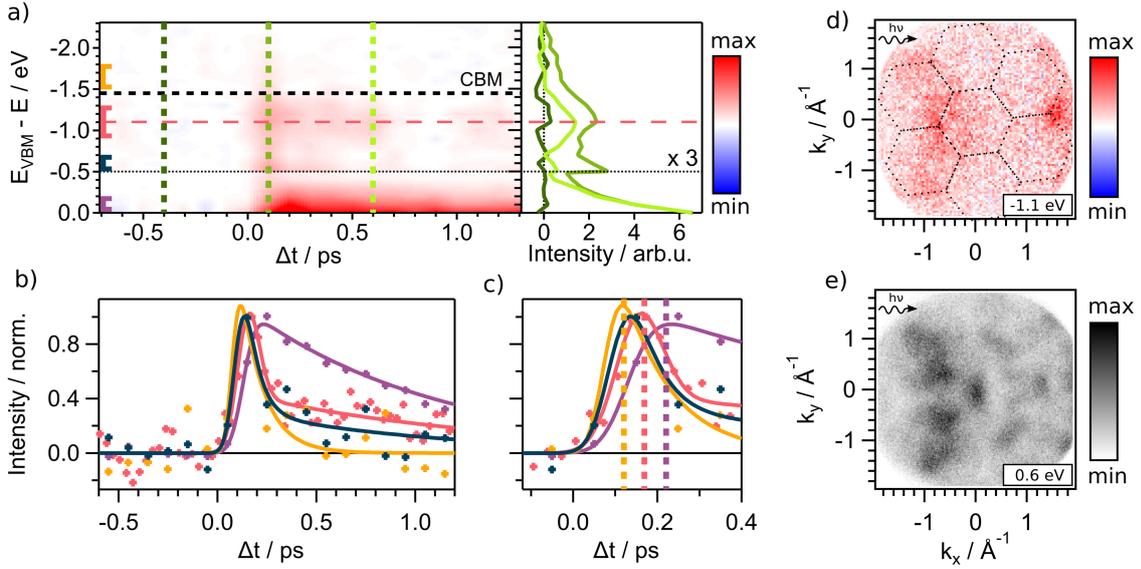

**Figure 3. Visualization and analysis of dynamics for hν$_2$=2.4eV excitation.** (a) Angle-integrated transient photoemission intensity, interpolated with a bilinear algorithm. For clarity, the signal from the unoccupied states 0.5 eV above the VBM is multiplied by factor 3. The right side of the plot shows the differential EDC for 3 different time steps, revealing a gaussian feature centered around -1.1 eV (marked by the red dotted line). The differently colored brackets on the left axis denote specific energy windows, the data from which are integrated to produce the normalized traced shown in (b). (c) shows a close up of the data around the peaks of the transient signal. (d) Differential momentum map at -1.1 eV, captured 130 fs after the initial excitation (t$_0$) at the peak of the transient signal. (e) Momentum pattern of the occupied bands of FePS3 at 0.6 eV below the VBM, underscoring the similarity between momentum patterns of occupied and unoccupied states.

In order to understand the states involved in the dynamics discussed above, we proceed as in the previous section and disclose the momentum-resolved intensity distributions at -1.1eV, which is the center of the energy region where the most interesting behavior emerges. This can be seen especially by the differential EDC curves that are shown on the right side of **Figure 3a** for three different time steps (corresponding to the vertical, dashed green lines on the left side). They reveal a long-lived Gaussian-like feature centered around -1.1 eV (indicated by the red dashed line). **Figure 3d** shows the differential momentum map captured at this energy 130fs after optical excitation. This map displays a distinct yet somewhat diffused k-dependent pattern, with peak intensities bordering the first Brillouin zone. When juxtaposing this pattern from the unoccupied states with momentum maps from occupied states around 0.6 eV below the VBM, there are evident resemblances to the occupied bands of $FePS_3$, as shown in **Figure 3e**. In this case, however, the resemblance is not as identical as observed in experiments with below-band gap excitation. In the next section, we will use this information to provide a complete picture about the observed transient dynamics. As we further approach the VBM, the transient signal shown by the purple curve, representing a 300meV energy window around the VBM, reverts to a single exponential decay, albeit with a longer decay time of $\tau_{purple}$ = 973 ± 66 fs. This signal is contributed to thermal broadening due to the above band gap excitation (More details in **Supplementary Fig. S2**).

As photoluminescence measurements suggest that above band gap excitation should also excite lower-energy d-d transitions[23], we further examine the subtle transient signal between 0.5 and 0.6 eV above the VBM (dark blue region) to uncover the presence of the $^5T_{2g} \rightarrow ^5E_g$ transition reported in the previous section. The transient trace shown with a dark blue curve in **Figure 3b** indeed contains two decay components, differing from the signal near -1.1 eV BE (red curve). Fitting this signal with a bi-exponential decay indicates that the slower decaying component can be ascribed to the signal from the VBM dynamics, while the faster component has a decay constant $\tau_{blue,2.4\ eV}$ = 66 ± 128 fs, comparable to the one observed from the $^5T_{2g} \rightarrow ^5E_g$ transition with $hv_1$ = 1.2 eV (more details in **Supplementary Figure S4**). A comparison of the peak position of the different extracted transients in **Figure 3c** shows a clear delay between the different signals. In reference to the signal with the fastest rising time, here the yellow curve of the excitation above the CBM, the maximum of the red curve lags the yellow transient by around 49 fs, while the purple transient is delayed by around 116 fs.

## 2.3 Localized model for the d-d excitations

To interpret the experimental findings, we start again from our assignment of the observed differential momentum map arising after excitation with 1.2eV photons to the spin-allowed $^5T_{2g} \rightarrow ^5E_g$ transition. This assignment can be rationalized by resorting to a simplified description for strongly correlated materials[24] based on the exact diagonalization of all local 3d charge configurations (see Methods). In this framework, the Coulomb repulsion U and the Hund's rule exchange J are entering as Racah parameters [24]. The 1.2eV pump pulse is theorized to induce a spin-allowed d-d transition, essentially transferring an electron from a doubly occupied $t_{2g}$ orbital to a singly occupied $e_g$ orbital, as schematically depicted in **Figure 4a**. The known excitation energy $\Delta E = 1.0786$ eV determines the crystal electric field splitting $\Delta_{CEF}$ of these orbitals[16]. This allows us to calculate the local Fe occupational (or lesser) spectral functions relevant for ARPES (details in **Supplementary section 7 and 8**). In static ARPES experiments, electrons are photoemitted from the Fe ground state configuration ($^5T_{2g}$). Conversely, trARPES experiments, particularly following a 1.2eV pump pulse, involve photoemission from the excited $^5E_g$ spin multiplet, resulting in a $3d^5$ configuration. This distinction is mirrored in the equilibrium and non-equilibrium atomic spectra, as depicted in **Figure 4b**. Crucially, the non-equilibrium spectra (shifted by the excitation energy $\Delta E$) and the equilibrium spectra exhibit several identical peaks, with slight differences in spectral weights. Since the atomic Green's functions become the starting point for the lattice spectral functions[25], and the hybridization between the orbitals remains unaltered by the laser pulse in leading order, the striking similarities between the non-equilibrium momentum maps depicted in **Figures 2c and 2d** with the equilibrium maps in **Figures 2e and 2f** can be well understood within this local Fe 3d picture.

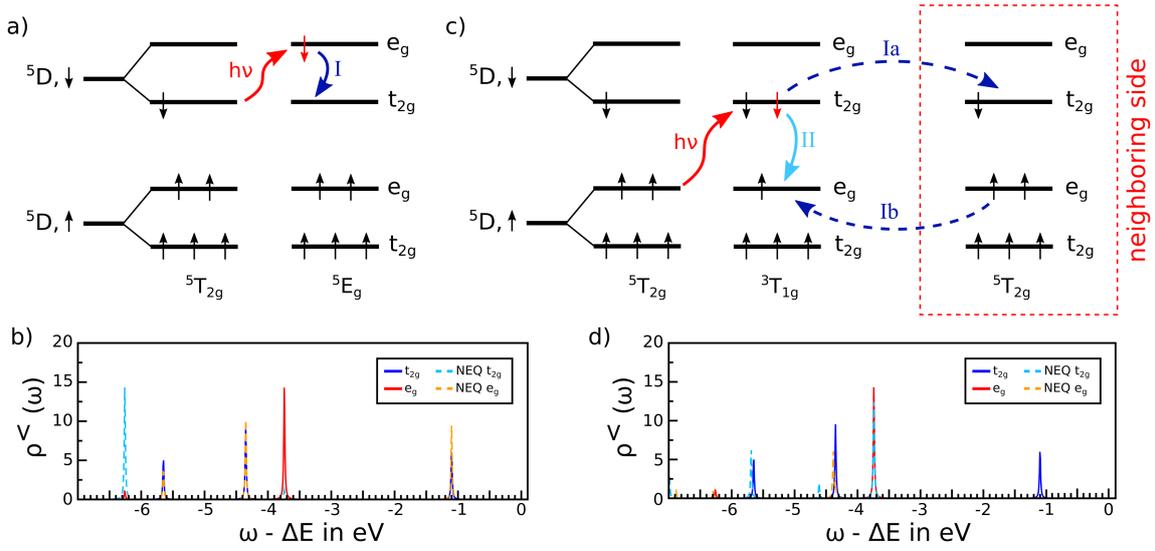

**Figure 4. Schematic representation of the multiplet state of the ground states and excited states and the corresponding calculations of the lesser Green function.** (a) Changes in the $Fe^{2+}$ ion multiplet state after exciting the energetically lowest lying d-d transition with $h\nu_1 = 1.2$ eV. This is the only transition possible without involving a spin flip due to a spin-conserving transition from the spin down $t_{2g}$ orbital into the $e_g$ orbital (shown by the red arrow). (b) The equilibrium spectra of the lesser Green function for the $t_{2g}$ (blue) and $e_g$ orbitals (red) as well as the non-equilibrium (NEQ) calculations depicted in dashed lines with the corresponding colors, shifted by the excitation energy of $\Delta E = 1.0786$ eV. (c) The 2nd lowest d-d transition involves a spin flip with an electron from the spin-up $t_{2g}$ orbital now occupying the spin down $t_{2g}$ orbital. This excitation can either propagate via virtual hopping to a neighboring side (Ia and Ib) or relax via SOC including a spin-flip transition (II). The corresponding equilibrium and NEQ spectra (shifted by the excitation energy of $\Delta E = 1.74$ eV) are shown in (d) and reveal a slight shift when comparing the different orbital peaks with a complete absence of the lowest excitation peak for the NEQ spectrum.

This approach allows us to interpret also the data obtained by above-band gap excitation. As mentioned before, the transient momentum maps measured at -1.1eV display a pronounced similarity to the momentum maps of the occupied states at 0.6eV. Given that this feature is too low in energy to be related to a population of the CBM, we ascribe the transient state observed at -1.1eV to the spin-forbidden $^5T_{2g} \rightarrow {}^3T_{1g}$ transition of the $Fe^{2+}$ multiplet, characterized by an energy of 1.7eV – the energy separation between the momentum maps of occupied and unoccupied states. **Figures 4d** presents a comparative analysis between the corresponding equilibrium and non-equilibrium atomic spectra, adjusted by the excitation energy. In this case, the spectra exhibit common features, though the differences are more pronounced compared to the previous case. This observation is consistent with the fact that the momentum maps in **Figures 3d and 3e** do not completely mirror each other.

Upon exciting $FePS_3$ with 2.4eV photons, our analysis reveals an initial population of the conduction band, which then depopulates with a time constant of $\tau_{yellow} = 115 \pm 100$ fs. The depopulation of the conduction band is followed by the excitation of the spin-forbidden $^5T_{2g} \rightarrow {}^3T_{1g}$ transition within 48 fs. Additionally, at 0.6eV above the VBM, the transient dynamics reveals the presence of the $^5T_{2g} \rightarrow {}^5E_g$ transition, that is also excited within 22 fs depopulation of the CBM. Closer to the VBM, the observed dynamics can be attributed to thermal broadening, with a thermalization time constant of $\tau_{purple} = 973 \pm 66$ fs.

The spin-forbidden $^5T_{2g} \rightarrow {}^3T_{1g}$ transition undergoes a double exponential decay, indicating the presence of two independent decay channels in the relaxation of the $^5T_{2g} \rightarrow {}^3T_{1g}$ transition back to the ground state. These two pathways have decay constants of $\tau_{red,1} = 46 \pm 34$ fs and $\tau_{red,2} = 1273 \pm 268$ fs. We associate them respectively to a virtual hopping between neighboring sides (exchange mechanism) and a spin-orbit coupling (SOC) mediated spin-flip process. **Figures 4c** visually portrays the excitation associated with the $^5T_{2g} \rightarrow {}^3T_{1g}$ transition and its relaxation back to the ground state.

To rationalize the timescale related to the SOC mediated spin-flip process discussed above, we carried out first-principles calculations that incorporate relativistic effects (detailed in Methods section). Starting from the electronic band structure of $FePS_3$ (**Supplementary Figure S9**), we determined the SOC constant for $Fe^{2+}$ to be $\lambda_{SOC} = -113.24$ cm$^{-1}$ (14.04 meV), primarily originating from the $Fe^{2+}$ d orbitals. The negative value of $\lambda_{SOC}$ reflects the more than half-filled occupied shell of $Fe^{2+}$ (with further details in **Supplementary Table S1**). By converting $\lambda_{SOC}$ in time ($\tau_{SOC}$ = h/E $\cong$ 300 fs), we find a reasonable correspondence with the experimentally determined spin-flip timescale $\tau_{red,2}$ = 1273 ± 268 fs. Additionally, we observe that in NiPS$_3$ the value of $\lambda_{SOC}$ = -280 cm$^{-1}$ (34.7 meV) was deemed to be sufficient to mix the $^1E_g$ and $^3T_{2g}$ excited states[16], thereby enabling the spin-forbidden $^3A_{2g}$ → $^1E_g$ transition. Considering that in NiPS$_3$ the energy separation between the SOC-mixed states is 7900 cm$^{-1}$ (0.98 eV), it is plausible that a similar SOC-mediated mixing of the $^5E_g$ and $^3T_{1g}$ excited states could be as well a viable source for spin relaxation in FePS$_3$. This supposition is bolstered by the fact that the energy difference in FePS$_3$ (0.72 meV) is smaller than that in NiPS$_3$ [16]

## 3. Conclusion

In summary, we have successfully identified two different d-d transitions in the antiferromagnetic van der Waals semiconductor FePS$_3$. By resonantly and non-resonantly exciting both states we were able to extract their lifetimes and identify the associated relaxation processes. Furthermore, we were able to assign specific momentum-dependent characteristics to both intra-ionic multiplet excitations in agreement with a theoretical model that adeptly encapsulates the primary microscopic features. We thus conclude that trARPES, when performed with a momentum microscope, can indeed capture the momentum-space signature of d-d transitions. This work extends the formidable suite of trARPES applications, complementing its established proficiency in identifying quasiparticles, such as excitons in molecular compounds[26,27] and light-induced quantum phenomena in 2D materials[28–30], with the ability to probe d-d transition dynamics.

## 4. Experimental Section/Methods

### 4.1 Sample preparation for photoemission spectroscopy

The samples were produced via CVD and commercially bought from HQ Graphene. The crystals are glued with a conductive silver epoxy on copper sample holders and were exfoliated with commercial scotch tape in situ at a base pressure in the low $10^{-8}$ mbar regime.

### 4.2 Femtosecond momentum microscopy

The experimental set-up consists of two major parts, namely, a photoemission electron microscope for ARPES experiments (KREIOS, Specs GmbH) detailed and benchmarked in Ref. [31] and a home-built system for the generation of fs-XUV radiation. In the following, we briefly describe the experimental set-up.

To track the dynamics of the localized d-d excitations between crystal-field split 3d states of the $Fe^{2+}$ multiplet in $FePS_3$, we use a pump–probe scheme. First, d-d transitions are excited (either directly or indirectly) with light pulses of photon energy 1.2eV and 2.4eV. Subsequently, the femto- to picosecond evolution of the excited $Fe^{2+}$ multiplet dynamics is probed with an XUV light pulse (21.7 eV, p-polarized), which photoemits the electron contribution of the multiplet excitation into the detector. The cross-correlation of pump and probe pulse was estimated to be $49 \pm 18.9$ fs.

The generation of coherent XUV radiation is driven by a commercial fiber-laser (Carbide, Light Conversion) with a central wavelength of 1035 nm, a temporal resolution of 242 fs, and an average power of 50 W. The variable repetition rate is set to 600 kHz for the reported measurements giving a pulse energy of 83 mJ. First, the pulses are compressed to sub 50 fs [32] by spectral broadening with self-phase modulation in a $X^3$ nonlinear medium followed by dispersion compensation with two mirrors with -200 fs² GDD. Second, the compressed pulses are frequency-doubled in a BBO crystal and tightly focused into an Ar gas jet inside a vacuum chamber. Here, high-harmonic generation (HHG) is used to create XUV pulses that are filtered by a pair of grazing incidence plates followed by an Al filter. Last, the 9th harmonic at 21.7 eV is selected in a monochromator with multilayer mirrors and focused onto the $FePS_3$ crystal. For the pump pulse, the remaining fundamental frequency after the BBO crystal travels through a motor-controlled delay stage to finely adjust the time difference between pump and XUV-probe.

The strength of the KREIOS analyzer is the simultaneous measurement of the in-plane momenta of the photoelectrons within the full photoemission horizon. In the instrument, an immersion lens column generates an image of the lateral distribution of the photoelectrons in the sample plane ($x$,y) and an additional Fourier lens transforms it into an image of the photoelectron emission angles. The latter contains the momentum distribution of the photoelectrons $I(k_x, k_y)$ with high angular resolution. The photoelectrons are then filtered by the hemispherical-analyzer section of the instrument, which acquires momentum-resolved two-dimensional (2D) maps at a fixed, and selectable, kinetic energy (E, $k_x$, $k_y$). For this reason, the instrument is also often called a momentum microscope [33].

## 4.3 Computational Methods

We described the Fe(II) 3d shell by a five-orbital model. The full Hamiltonian comprises single-particle orbital terms and the spin-rotational invariant general Coulomb interaction, parameterized by the Hubbard energy U and the Hund's rule exchange interaction J.

The single-orbital energies include a crystal electric field splitting between the $t_{2g}$ and $e_g$ orbital determined by spectroscopic excitation energy $^5T_{2g}$ → $^5E_g$, $\Delta E = 1.076 eV$[16] and are adjusted such that the ground state contains six electrons. The local Mott excitation $U_{eff} = U-3J = 2.2$ eV is taken from a recent LDA+U approach to FePS$_3$[34]. The demand to reproduce the lowest experimentally observed $^5T_{2g}$ → $^3T_{1g}$ transition at $\Delta E = 1.796 eV$ determined the two Racah parameters to U = 4.15eV and J = 0.65eV. We used the total spin and the total charge as conserved quantum numbers to set up the block diagonal many-body Hamiltonian and determined its spectrum and eigenstates by exact diagonalization of each block. From the results, we calculated the spectrum of the lesser Green's functions describing the removal of an electron from a $t_{2g}$ or $e_g$ orbital orbital of the 3d shell, using either the projector onto the ground state manifold (equilibrium) or the projector onto one of the excited states (non-equilibrium) as density operator.

The electronic structure including spin-orbit coupling was calculated by means of DFT + U + SOC using the projector augmented wave (PAW) method as implemented in the Vienna Ab initio Simulation Package (VASP)[35]. We considered van der Waals (vdW) interactions between layers using the DFT-D3 method[36], and the plane-wave cutoff was set to 350eV. The structural optimizations of atomic coordinates were performed setting a force and energy convergence criteria of 0.001 eV/Å and $1.0 \times 10^{-6}$ eV, respectively. The Brillouin zone was sampled with a Γ-centered 4 x 4 x 4 k-point Monkhorst–Pack mesh[37]. To properly describe the strong correlation in the d orbitals of Fe we adopted a Hubbard $U_{eff} = 2.2 eV$, using the simplified version proposed by Dudarev et al [38]. A 1 x 1 x 2 monoclinic cell to account for the antiferromagnetic (AF) interlayer coupling was employed.

**Acknowledgements**

We acknowledge help from Evgeny Zhukov in recording the absorption spectra.

This work was supported by the Deutsche Forschungsgemeinschaft through the TRR 142/3, Project A08. We acknowledge support from the European Union's Horizon 2020 Research and Innovation Programme under Project SINFONIA, Grant 964396 and ERC-StG-101042680 2D-SMARTiES. Financial support from MERCUR "Kooperation" project: "Towards an UA Ruhr Ultrafast Laser Science Center" is also acknowledged. The Momentum Microscope has been financed by the Deutsche Forschungsgemeinschaft (DFG) through the project INST 212/409 and by the "Ministerium für Kultur und Wissenschaft des Landes Nordrhein-Westfalen". J.J.B. acknowledges the Generalitat Valenciana (CDEIGENT/2019/022) and A.M.R. thanks the Spanish MIU (Grant No FPU21/04195).